# Domain resource integration system

Wang Liang[1], Guo Yi-Ping[2], Fang Ming[3]

[1, 3] (Department of Control Science and Control Engineer, Huazhong University of Science and Technology, WuHan, 430074 P.R.China)

[2] (Library of Huazhong University of Science and Technology, WuHan 430074 P.R.China)

E-mail: wangliang_f@yahoo.com

Phone: +86-27-87553494

**Abstract** Domain Resource Integrated System (DRIS) is introduced in this paper. DRIS is a hierarchical distributed Internet information retrieval system. This system will solve some bottleneck problems such as long update interval, poor coverage in current web search system. DRIS will build the information retrieval infrastructure of Internet, but not a commercial search engine. The protocol series of DRIS are also detailed in this paper.

**1 introduction**

Now most people obtain information from Internet by search engines, which give us great convenience in our daily life. But for many people, finding the information they really want on the World-Wide Web is still a hit-and-miss affair. We can find some bottleneck problems in current search engines.

Because the Internet is a large dynamic system, current search engine can't continue to index close to the entire Web as it grows. The update interval of most pages database is almost one month. Till now, no a search engine can cover more than 50 percentage pages on Internet. These data has no obvious difference with the statistical data in 1998[1]. In fact, the performance in current search engine is even no better than those of six years ago. Web page is only one kind of information resource on Internet. Many video, PDF, picture and many other kinds of information resource also greatly increased in these years. Some search engines like Google also want to add these resources in their system. But only as a web page search engine, it has meet many problems in coverage and recency, etc. If it adds the other kinds of resources, these problems may become more serious.

Moreover, there have been hundred of different search engines [2]. Every of them will extract all of the pages from the web servers over and over again, which great increase the traffic load of Internet. Personal information is not considered in current search system. So there are always too many unrelated records in search results. At all, we need a better internet search engine solution.

**2 DRIS**

To solve these problems in current search system, a new system, domain resource integration system (DRIS), is proposed in a digital library project. DRIS was also a selective solution of IETF [3] to build the new internet information retrieval infrastructure.

**2.1 Origins of DRIS**

Firstly, the original idea of DRIS is introduced. Now more and more digital resources were introduced into library. Just in our library, there have been one hundred kinds of resources such as IEEE, ACM, many Chinese digital journals, etc. We can also get information from many public resources like web search engine. We may always feel very inconvenient to find the precise search results in hundreds of recorders of Google, but when we go to library, we will find it's just a beginning. You may have to search in tens of different information resources one by one. You should also be very familiar with the query rules of every database. Then you may get the comprehensive and precise information. It's a really difficult mission. Everyone hopes to obtain all kinds of information on Internet in one system. The search results of this system should also be precise, no many unrelated recorders.

When Internet became the main data source of library, many some problems of Internet also becomes the main research topic of digital library research. The Internet information retrieval problem is more serious in digital library.

To solve this problem, we need to complete two missions. First, we should build a system that can integrate all the resource on Internet. Current information Internet is still a fragmentized world. There was still no efficient connection among different resources. Second, after finishing this resources integration system, some search system will apply it as data source and provide union search service for user. Building a usable internet information infrastructure and providing more efficient search service become the main goal of digital library research.

**2.2 basic ideas of DRIS**

To finish the two missions mentioned above, we completely divide the Internet search engine into two parts: Internet information retrieval infrastructure and personal search system. This is just the main difference between current search engine and DRIS. DRIS will be treated as the public information retrieval infrastructure, which will integrate all kinds of resources on Internet. The personal engine system can organize, rank, filtrate the search results according your personal information. DRIS will be its data source. In this method, you can get more precise information in.

So the basic idea of DRIS is that search should be the international function of Internet and everyone should have his own search engine.

**2.3 The architecture of DRIS**

The final aim of DRIS is to integrate all the resource on Internet. Now there have been billions of web pages, millions of special databases and many other kinds of information resources on Internet. Gathering all the resources in a system and building a mirror database of whole Internet may be an impossible mission. Even we can have an enough big storage system and a powerful processor system, the update interval and coverage of its data also can't be ensured. Centralized architecture is not appropriate to build Internet information retrieval system. But most current commercial search engines all apply it.

Normally distributed management is much more effective than centralized administration in a large-scale system. So we also adopt a hierarchical distributed architecture to manage all the information on Internet. By this means, the key issue is how to divide the Internet correctly. We found there has been an available division method on Internet, domain name system (DNS). DNS is a hierarchical distributed system. All the web sites on Internet are efficiently managed in this system. We also apply its basic idea to DRIS. The basic architecture of DRIS is as same as DNS, which is shown in Fig 1.

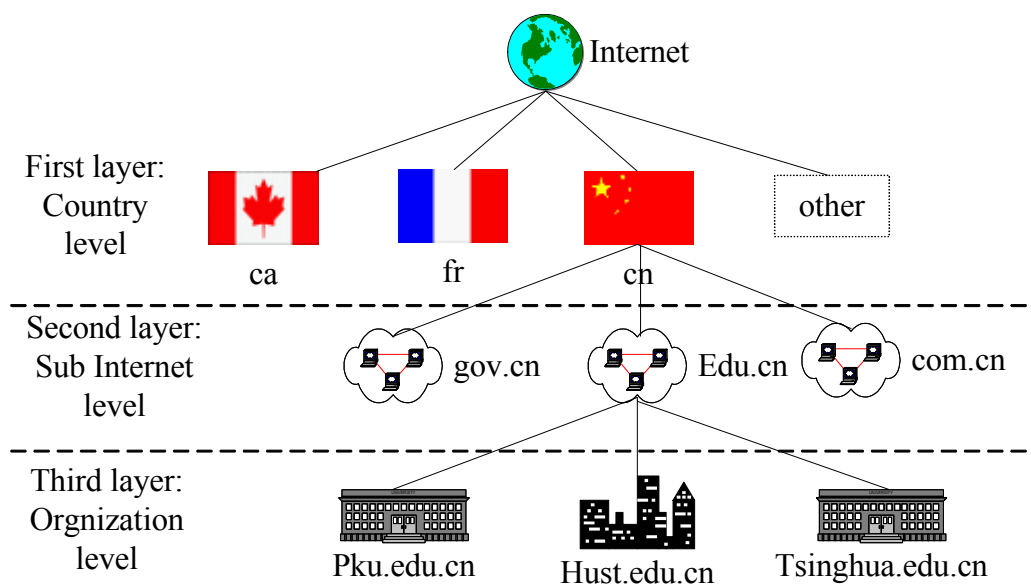

Fig the architecture of DRIS

There are three layers in DRIS. The third layer corresponds to some organization live university. The second is always the sub Internet of a country. The top layer corresponds to each country. The three layers of DRIS strictly correspond to different layers of DNS.

**2.4 search system in each layer**

To build an efficient internet information retrieval infrastructure, DRIS applies different kinds of search system architecture in different layers.

Three kinds of search systems are introduced as follows.

1 Centralized search system based on conventional database system. This system has its own data collecting mechanism, and all the data are stored and indexed in a database system. Although many web search engine provide service from thousands servers, their database all are the same and also belong to this kind of search system.

2 Search system based Metadata harvest system. When we need to integrate different kinds of information resource like video, pdf, web pages in a system, we can harvest the metadata from sub databases and build a union metadata database to provide the combined search function. Normally Metadata is much smaller than data itself. Some systems based on OAI like NSDL just apply this method.

3 Distributed search system[4]. Distributed information retrieval system has no his own actual record database. When receiving a query from a user, it will instantly obtain the records through the search interfaces provided by sub databases. A famous system is the infoBus system in Stanford digital library project [5].

How to select correct information retrieval architecture? There are two main characters to determine the architecture of an information retrieval system, the size and diversity of data source. Normally, with the increase of size and diversity of data source we can select conventional database system, metadata harvest system and distributed search system respectively.

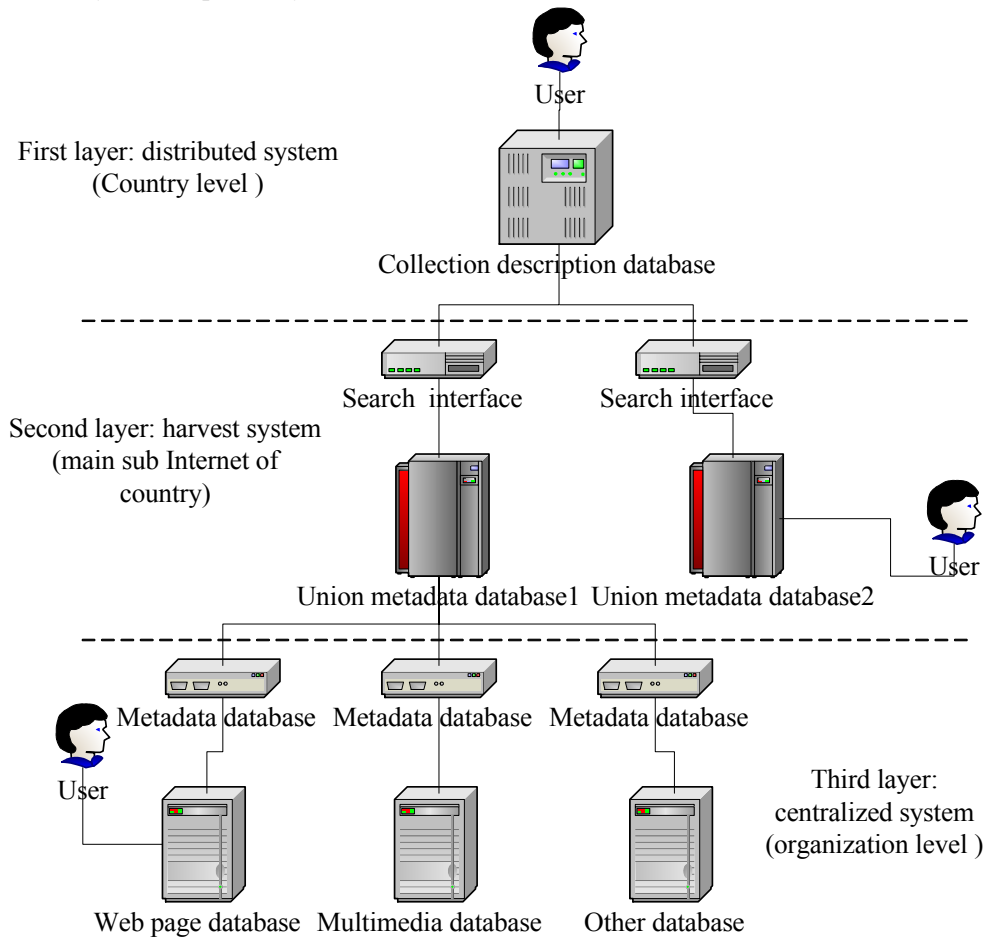

Fig2 function architecture of DRIS

We apply this principle in DRIS. We can find the three layers of DRIS nicely correspond with three kinds of search system. From the third layer to top layer, we can select conventional database system, metadata harvest system and distributed search system respectively. The function architecture of DRIS is shown in Fig2.

In the third layer, centralized database system is applied to build search system. Node in this layer is always limited in a university or a company, and the size of its data source is very small. For example, the number of web pages in a university normally is not more than one million, and most of current database systems can easily manage this kind of data source.

In the second layer, we adopt the metadata harvest system to build the information retrieval system. As the web search engine of node "Edu.cn", we will integrate the all the web pages in all universities in China. Merely harvesting the metadata from thousands of DRIS servers in the third layer will be more efficient than directly downloading all the web pages from millions of web servers. The other kinds of resources can also be integrated by this means.

In the first layer, Distributed information retrieval system is applied to integrate the information resource in the scope of a country. In this layer, only the collection description data of resources is stored in database of DRIS. All the resources will cooperate to finish a user's query under the directing of DRIS.

DRIS suitably applies three kinds of information retrieval system to build an efficient information retrieval infrastructure for Internet. The basic principle of DRIS could be described as follows: Third layer-Organization lever- conventional database system; Second layer- Sub Internet level – metadata system; First layer-Country level- distributed search system.

**2.5 management and application of DRIS**

The next question is how to organize this big system and how to use this system.

As an application software system, we normally use class tree of OO (Object Oriented) model to describe DRIS. The class tree of DRIS is shown in Fig 3.

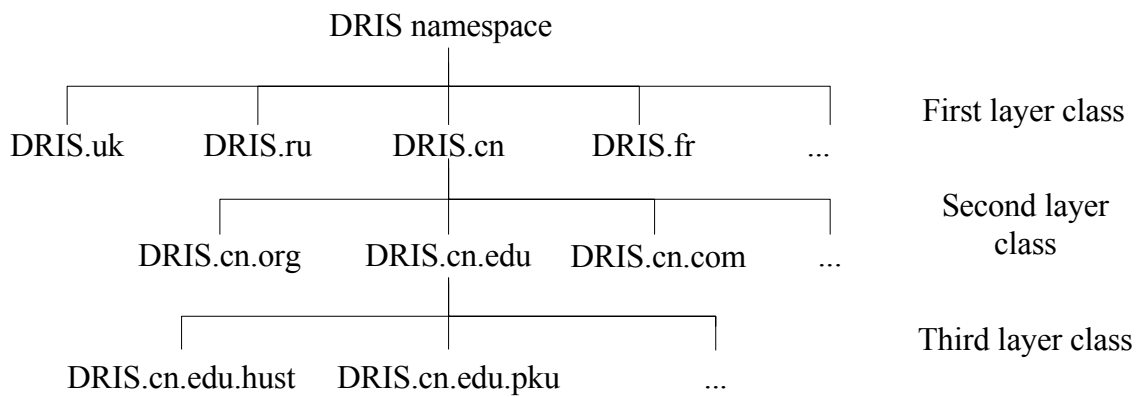

Fig3 a OO model of DRIS

All the nodes in the DRIS are arranged under the namespace "DRIS" and be treated as the child class of DRIS. These child classes and their function are realized in different DRIS servers. DRIS is distributed system, so we apply Webservice to organize and connect the whole DRIS. We define some basic rules for the management of DRIS.

1 All the nodes of DRIS will provide the search service in the form of standard Webservice.

2 All this service is organized by the inheriting relation, but the realization of this relation has some differences with that of conventional OO model. The node in lower layer can inherit the class in higher layer by citing the Webservice of node in high layer. But because DRIS is a distributed system, the node in higher layer is lies in another server and can't know its child class. So the higher node will have a special mechanism to index all its child class.

3 The Webservice is based on SOAP and provides the service by its URL. Every node of DRIS is an integrated search engine and provides its stand search Webservice. But for the end user, how can they find the appropriate search service quickly. For example, I want to find a university search Webservice, but what's its URL? So for the advantage of end user, we define a special rule for the location of DRIS server. If main class name of this Webservice is "DRIS. Class name", its Webservice should provide the service through URL of "http://DRIS.Class name". It's also the domain name of corresponding DRIS server. Because Class name of each node of DRIS corresponds to the each Domain name, this rule can also be described as that all search Webservice should provide the service through URL of "DRIS. Domain name" and the main class name of this Webservice will be "DRIS. Reversed domain name". For example, the domain name of our university is "hust.edu.cn". Then our DRIS server will provide the search service through link of "http://DRIS.hust.edu.cn" and its main class name is just "DRIS.cn.edu.hust ".

So if the DRIS is implemented, all Domain names "DRIS. Domain name" should be left to DRIS servers. In DRIS, Domain name is used not only in navigation but also as the identification of aggregate of information resources. DRIS could be regarded as the update of DNS.

This is just the basic rule of DRIS. Detailed content will be defined in the protocols of DRIS.

The class tree of DRIS server in our university is shown in Fig4. The top class "DRIS.cn.edu.hust" will provide a union search Webservice of all the resource in our school. Class "DRIS.cn.edu.hust.webpage" will provide the web pages search service in the scope of our university. Class "DRIS.cn.edu.hust.ftp" will provide the ftp search service. The other kinds of resources also issue their standard search service in DRIS server.

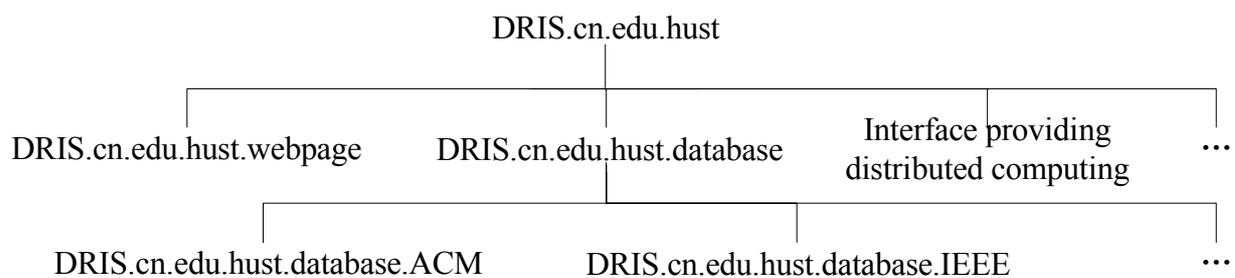

Fig 4 a DRIS node in the third layer

Another question is how can we get information from DRIS? In DRIS, every node is an integrated search engine and can provide standard search Webservice. Any intelligent search systems can apply DRIS as their data source and provide high quality of personal search service. We can get what we really want from our own search engine, but not from public search system.

**3 DRIS protocols**

DRIS is based on public protocol. To make the whole system more clear, we divide the DRIS protocol series into five parts. In fact, every protocol is an integrated system and solves a special problem.

This is the five parts of DIRS:

1 Standard distributed search system. It defines the platform-independent search interface and a collection description standard for heterogeneous information resources.

2 Standard metadata harvest system. A protocol based some available opening standard like OAI will be proposed. It will define a standard metadata that can be compatible with most database system.

3 Standard public web pages search system. A web search engine comprises three parts, Page-fetching mechanism, page database system, search interface. Corresponding protocol will also be made up of these three parts. There are many kinds of digital resources. As long as they can provide the standard distributed search interface or comply with the metadata harvest format, they can all be brought into DRIS. But web pages search system is a large distributed dynamic system. It will strictly comply the basic principle of DRIS {(organization level-download and index pages)-(sub internet level-harvest metadata)-(country level-distributed system) to build an efficient web search system.

4 Whole DRIS. It includes its whole architecture, the relation between different nodes, etc.

5 DRIS and IPV6.IPV6 will be the most distinct feather of next generation Internet.IPV6 is still in improving and any technology that can benefit the Internet all can be added to the IPV6 system. Since the searching is the main service of most user of Internet and this service is not so satisfied to us in current Internet, why not take this request into account when build the new Internet. For example, in IPV6, all kinds of data flows are assigned a priority, and then Internet can guarantee a high priority to the data flow of DRIS.

**4 the implementation of DRIS**

Although DRIS gives us an excellent and promising solution for the new Internet search system, this can't ensure the establishment of DRIS. We should found some urgent request for this system. We have built some experimental systems in CERNET (China education and research network, including all the universities in China).

The nodes in the third layer DRIS in CERNET correspond to different universities in China. We can easily found the request to build such system. Every of us may feel very confused when searching in hundreds of different digital resource in library one by one. We need a union search system that can integrate all the information in school network, including digital resources in library, web servers in school, etc. It's just the request in the third layer. Then sharing the resources in different universities may bring the request to build the DRIS server in the second layer, which will integrate all the resource on CERNET.

To carrying out DRIS in larger scale, the implementation IPV6 may be a good chance for DRIS. Number of IP address, security, QoS and other problems are all improved in IPV6, but as one of most important services on Internet, its problem may be more serious and urgent. On the other hand, IPV6 also need a "killing application" for its own application.

**5 Conclusions**

The distinct architecture of DRIS can ensure the update interval and coverage of search system. Dividing the search engine into two parts will give the users higher quality of search service. DRIS is public Internet information retrieval system .The application of Webservice can provide excellent data source. These will give great advantage to develop new intelligent search systems. At all, a public Internet information infrastructure is very necessary for the further development of Internet.